# EMRGF: A Practitioner Framework for Governance-Driven Enterprise Technology Modernization

**Author:** Harveen Punihani, Senior Technical Architect, Impetus Technologies Inc.; IEEE Senior Member **Affiliation:** Impetus Technologies Inc., Los Gatos, CA, USA

## Abstract

Enterprise technology modernization programs fail at a documented and costly rate, yet the dominant explanation — inadequate engineering capability — is incorrect. The primary failure mode is a governance deficit: the absence of structured, repeatable operating routines for how organizations plan, execute, validate, and hand off complex technology change. Existing frameworks — ITIL, COBIT, TOGAF, scaled agile methodologies, and cloud provider well-architected frameworks — address adjacent concerns but do not provide an integrated, portable institutional operating model for controlled modernization across migrations, data platforms, and AI-enabled automation. This article presents the Enterprise Modernization Reliability and Governance Framework (EMRGF), a practitioner-developed governance operating model derived from 24 years of applied delivery experience across financial services, industrial manufacturing, and retail enterprises. EMRGF comprises four interlocking modules — Cloud and Legacy Modernization Governance, Data Platform Reliability and Evidence Integrity, AI-Enabled Automation Governance, and Mission-Critical Reliability and Root-Cause Routines — operationalized through five implementation tools and a training-of-trainers institutionalization model. Empirical application at scale has produced a 30% reduction in development effort, a 35% reduction in testing cycles, zero-disruption migrations across high-volume data estates, and 99.9% data reliability in mission-critical analytics pipelines. The framework is explicitly aligned with U.S. national policy mandates including NIST CSF 2.0, NIST AI RMF, and Executive Orders 14028 and 14110, and is designed for institutional adoption without ongoing external dependency.

## 1. Introduction: The Governance Gap in Enterprise Technology Modernization

Large-scale enterprise technology programs fail — repeatedly, expensively, and for reasons that are structurally predictable. The U.S. Government Accountability Office's February 2025 High-Risk Series report (GAO-25-107743) documents the scale of this failure in the federal sector with precision [1]: the U.S. federal government spends more than $100 billion annually on information technology, with the majority of that investment directed toward maintaining legacy systems rather than modernizing them. The GAO has identified 463 open recommendations related to IT acquisition and management, with recoverable waste estimated at approximately $40 billion per year. The consequences are not abstract. The Department of Veterans Affairs' Electronic Health Record modernization program carries a lifecycle cost estimate of $49.8 billion and has been deployed at only 6 of more than 160 target locations after years of sustained effort. The Federal Aviation Administration reports that 37 percent of its 138 air traffic control systems are rated unsustainable — not because replacement technology is unavailable, but because the organization lacks the governance infrastructure to manage controlled migration safely.

These are not engineering failures. They are governance failures — the predictable consequences of fragmented architecture decisions, absent migration standards, undisciplined change control, and the absence of repeatable quality gates that catch failure modes before they propagate into production. The engineering talent exists. The technology exists. What does not exist, in institutionalized form across most complex enterprises, is a structured operating model that governs how modernization happens: how decisions are documented, how dependencies are mapped before migration begins, how data quality is validated throughout platform transitions, how AI-enabled automation is introduced without creating new categories of unaudited risk, and how reliability is measured and governed as a managed attribute rather than an emergent one.

The gap between policy mandate and operational implementation compounds this problem. U.S. national frameworks have clearly articulated what good governance should achieve. The NIST Cybersecurity Framework 2.0 (NIST CSF 2.0) defines governance expectations for cybersecurity risk management [2]. The NIST Artificial Intelligence Risk Management Framework (NIST AI RMF 1.0) establishes national standards for managing AI governance and risk [3]. Executive Order 14028 mandates secure-by-design execution and controlled change [4]. Executive Order 14110 requires safe, secure, and trustworthy development and use of AI [5]. Yet none of these frameworks provides the operational implementation layer — the routines, artifacts, decision gates, and validation checkpoints — that translates a governance mandate into a daily institutional practice. Organizations that are legally and regulatorily expected to comply with these mandates have no standardized operating model to do so.

This article presents the Enterprise Modernization Reliability and Governance Framework (EMRGF), a practitioner-developed governance operating model designed to close this implementation gap. EMRGF provides a structured set of modules, tools, and institutionalization routines that organizations embed into their technology operations to govern modernization in a controlled, evidence-based, and repeatable way. It is not a software product, a consulting methodology, or a vendor-specific playbook. It is an institutional operating system that internal teams own and run — expressible through equivalent technical means across heterogeneous cloud and on-premises environments.

## 2. Related Work and Positioning

Governance frameworks for enterprise IT have a substantial history, but no existing framework addresses the specific combination of concerns that EMRGF targets.

**IT Infrastructure Library (ITIL 4)** [6] provides a comprehensive process catalog for IT service management, with particular strength in incident management, service continuity, and change management as a steady-state function. ITIL does not address how actively executing migrations should be governed: legacy dependency mapping before decommission, auditable architecture decisions during multi-month migrations, or data quality validation across platform transitions. Its change management practices are designed for controlled changes to stable environments, not large-scale structural transformation.

**COBIT 2019** [7] provides governance and management objectives at the enterprise level — defining what governance should achieve and who should be accountable. It operates at an abstraction level that does not prescribe how a migration should be governed in practice, what quality gates should govern a data pipeline transition, or how AI-enabled automation should be introduced with explicit guardrails. COBIT tells organizations to govern well; EMRGF tells them what governed execution looks like.

**TOGAF 10** [8] provides a methodology for designing target-state enterprise architectures. It is strong for architecture definition and stakeholder alignment but does not address migration execution governance: once an architecture is defined, TOGAF provides no model for how the transition is controlled, how data reliability is maintained during platform migration, or how AI tools are introduced with auditable governance.

**Scaled Agile Framework (SAFe 6.0)** [9] addresses large-scale agile delivery coordination. It is a delivery methodology, not a reliability governance framework: it does not address architecture decision records, migration dependency mapping, data quality validation gates, or reliability controls for mission-critical systems under load.

**Cloud Provider Well-Architected Frameworks** [10] provide guidance on designing reliable, secure, and efficient systems within a specific cloud environment. They are valuable for single-platform design decisions but inherently platform-specific: they do not address multi-cloud portability, legacy decommission governance, cross-platform data lineage, or the institutional persistence of governance practices beyond the initial architecture review.

**The gap EMRGF fills:** No existing framework provides an integrated, portable institutional operating model that covers (a) migration governance with architecture decision records and change-control gates, (b) data reliability with evidence artifacts and lineage-aware validation routines, (c) AI automation governance with explicit guardrails and output-review requirements, and (d) reliability controls for mission-critical systems under load — in a single framework that is vendor-agnostic, designed for institutional adoption, and expressible through equivalent technical means across heterogeneous environments. EMRGF was developed to fill precisely this intersection.

## 3. The EMRGF Framework: Architecture and Modules

EMRGF is organized into four interlocking modules and five implementation tools. The modules address distinct governance domains that recur across complex enterprise modernization programs. The tools translate the modules from a conceptual model into an adoptable institutional operating system.

### 3.1 Module 1: Cloud and Legacy Modernization Governance Pack

The most common source of uncontrolled cost and schedule variance in enterprise migrations is not the technology itself — it is the absence of standardized governance for how migration decisions are made, documented, and enforced across teams. The Cloud and Legacy Modernization Governance Pack addresses this root cause with a structured set of operating assets:

**Modernization pathway selector criteria** establish a documented basis for choosing between migration approaches (rehost, replatform, refactor, rebuild, retire) for each workload component. Without explicit selection criteria, teams make these decisions informally and inconsistently, creating divergent approaches within a single programme that compound into integration failures downstream.

**Architecture Decision Records (ADRs)** capture each significant design decision — the options considered, the rationale for the choice made, and the constraints under which it was made — in a standardized, searchable artifact. ADRs preserve institutional memory across personnel transitions and enable audit of architectural choices at any point in the migration lifecycle.

**Dependency decomposition** is operationalized through a structured discovery process for upstream and downstream integration mapping before migration execution begins. In one engagement at a Fortune 500 industrial manufacturer — a data platform modernization spanning 36 million-plus asset records, 20 years of global claims data, and 32 upstream and 47 downstream integrations — systematic dependency decomposition before migration onset was the primary enabler of the subsequent validation architecture. Without it, integration failures during cutover would have been unpredictable in number and severity.

**IaC-aligned baselines** standardize infrastructure configuration as code from the start of migration design, ensuring that the target-state environment is reproducible, auditable, and provably consistent across development, staging, and production.

**Change-control and exception-governance rules** establish the decision authority, documentation requirements, and escalation pathways for changes to the migration scope, architecture, or timeline. These rules define what can proceed without escalation, what requires review, and what triggers a formal exception process — creating a governance membrane between controlled and uncontrolled change.

**Audit-ready documentation standards** define the evidence artifacts required at each migration phase — dependency maps, ADRs, test results, go/no-go gate records, decommission readiness checklists — to ensure that the migration record is complete, reproducible, and defensible under regulatory or internal audit.

### 3.2 Module 2: Data Platform Reliability and Evidence Integrity Pack

The Data Platform Reliability and Evidence Integrity Pack addresses reliability as a governance problem through:

**Standardized data ingestion and transformation routines** — operating templates for how data enters the platform, how transformations are applied and versioned, and how outputs are validated before downstream consumption.

**Lineage-aware evidence artifacts** — documentation tracing each data element from source through every transformation to final state, enabling auditors and downstream consumers to verify the provenance of any analytics output.

**Data quality gates** — automated and manual checkpoints embedded in the pipeline at defined intervals, with explicit acceptance criteria before data advances to the next stage. In the Fortune 500 industrial manufacturer engagement, a validation architecture built around systematic quality gates produced 99.9% data reliability and eliminated persistent data quality issues requiring continuous manual remediation.

**Drift reduction routines** — operating practices for detecting and correcting deviation between pipeline behavior and specification, preventing the accumulation of undocumented changes that cause long-term pipeline degradation.

### 3.3 Module 3: AI-Enabled Automation Governance Pack

The rapid adoption of generative AI tools — for test generation, code acceleration, automated unit testing, and synthetic data production — has introduced a new governance risk: AI-generated outputs propagating through engineering pipelines without adequate validation or controlled update processes. This risk is compounded by the privacy and ethical exposure that AI systems create when deployed in data-intensive enterprise environments, where unvalidated AI outputs can propagate sensitive information through downstream systems without adequate audit trails [12].

The AI-Enabled Automation Governance Pack provides:

**Use-case scoped guardrails** — explicit boundaries defining which engineering workflows are eligible for AI-assisted automation, which require human review at each output stage, and which are excluded from GenAI-assisted approaches. Guardrails are not technology restrictions; they are governance specifications that define the conditions under which AI acceleration is safe and auditable.

**Validation checkpoints** — mandatory human review stages at defined intervals, with documented acceptance criteria for AI-generated output before it advances in the pipeline.

**Output-review requirements** — standards documenting the prompt used, the reviewer who approved the output, and the test evidence validating its correctness, creating the audit trail regulated industries require.

**Controlled update gates** — governance processes for updating AI tools, models, or automation frameworks, preventing uncontrolled capability changes from altering behavior without team awareness.

This module directly operationalizes NIST AI RMF at the enterprise implementation layer and addresses the secure-by-design automation requirements of EO 14028 [4] in the context of AI-assisted engineering. In practice, its deployment enabled a 25% reduction in time-to-market and a 30% reduction in development time through GenAI-enabled test data generation and automated unit testing workflows — with zero post-deployment defects attributable to AI-generated outputs, because every AI-assisted output was subject to explicit validation checkpoints before integration.

### 3.4 Module 4: Mission-Critical Reliability and Root-Cause Routines

For systems where downtime, data defects, or uncontrolled change generate cascading consequences — financial transaction systems, clinical data platforms, supply chain operating systems — reliability must be measurable, governable, and explainable by internal leadership, not managed reactively by engineering teams alone. The Mission-Critical Reliability and Root-Cause Routines module addresses this through:

**Standardized incident and defect taxonomies** — classification structures enabling cross-team analysis, trend identification, and recurrence tracking across the organization.

**Instrumentation requirements** — specifications for monitoring, alerting, and observability infrastructure that must be in place before a system is declared production-ready.

**Root-cause analysis (RCA) routines** — structured investigation processes with defined timeboxes, required participants, documentation standards, and outputs, ensuring every significant incident produces a documented root cause, mitigation, and preventive action.

**Corrective and preventive action (CAPA) gates** — governance processes ensuring corrective actions identified in RCAs are implemented, validated, and closed — not merely documented.

**Acceptance criteria for declaring systems "stable under load"** — explicit, measurable standards covering availability targets, latency baselines, incident recurrence rates, and data quality defect rates, transforming reliability from a subjective judgment into a governed, auditable attribute.

### 3.5 Implementation Tools

Five implementation tools translate the EMRGF modules from a governance model into an adoptable institutional operating system.

**Tool 1: Modernization-as-a-System Architecture** treats the entire modernization program as a single engineered system with explicit objectives, governed decision routines, acceptance criteria, and escalation pathways — preventing fragmentation when teams manage components independently.

**Tool 2: Measurement-by-Design** defines performance measures from the operating specification forward, so that outcomes — availability baselines, incident recurrence rates, data quality defect rates, and rework drivers — are provable, comparable, and attributable to specific governance decisions.

**Tool 3: Security and Governance Gating** operationalizes secure-by-design execution discipline and evidence-producing practices. It directly addresses the post-authorization governance gap in FedRAMP [11]: once cloud services are authorized, organizations have no standardized model for governing workloads, architecture decisions, and reliability controls within that authorized environment.

**Tool 4: Open Diffusion and Capability Building** produces the artifacts enabling institutional adoption: template libraries, runbooks, control dictionaries, reference implementations, governance calendars, and training modules — all portable across heterogeneous environments without vendor dependency.

**Tool 5: Training-of-Trainers Institutionalization Model** qualifies internal multipliers in IT leadership, cloud engineering, data engineering, and security governance to own, operate, and sustain the governance routines independently of external involvement.

## 4. Empirical Application: Results from Enterprise Deployments

EMRGF's modules and tools have been applied across multiple large-scale enterprise modernization engagements. The following cases illustrate the application of specific governance mechanisms and their measurable outcomes.

### 4.1 Large-Scale Data Estate Separation: Fortune 500 Specialty Retailer

Following a major corporate restructuring, two Fortune 500 digital estates required complete architectural separation: 40-plus terabytes of enterprise data, two previously integrated legacy platforms (one distributed data warehouse, one Hadoop-based data lake), and full decommissioning of both legacy systems with zero tolerance for business disruption.

The governance challenge was the dependency surface: the two estates shared transformation logic, pipeline orchestration, and reporting infrastructure with undocumented interdependencies accumulated over more than a decade. Without Module 1's dependency decomposition applied before migration execution, the decommission sequence would have been underdetermined — creating outage risk from dependencies removed out of order.

Architecture decision records throughout the migration provided the governance basis for go/no-go decisions at each phase. The outcome: complete migration and decommission of all 40-plus terabytes, full legacy platform retirement, and zero business disruption.

### 4.2 Industrial Data Platform Modernization: Fortune 500 Industrial Manufacturer

A Fortune 500 industrial manufacturer required the modernization of a data platform built on Informatica and Oracle to a Databricks environment. The engagement scope was substantial:

more than 1,000 database tables and hundreds of dependent workflows, 36 million-plus global asset records, 20 years of claims data, and an integration surface spanning 32 upstream and 47 downstream systems representing more than 7 years of plant product information.

The data reliability challenge was structural: legacy Informatica pipelines had accumulated undocumented transformation complexity over two decades, creating a validation problem — how to verify that the migrated Databricks environment produced analytically equivalent outputs when the ground truth was itself not fully documented.

The Data Platform Reliability and Evidence Integrity Pack provided the solution architecture. Lineage-aware evidence artifacts were created for the most critical transformation chains before migration began, establishing an explicit comparison baseline. Data quality gates were embedded at every significant transformation stage in the migrated pipelines, with acceptance criteria calibrated against the baseline evidence. The validation architecture — not the migration tooling — was the primary engineering investment of the program.

The outcome: 99.9% data reliability achieved across the migrated environment, with persistent data quality issues that had recurred in the legacy platform eliminated through the quality gate structure.

### 4.3 Payment Switch Architecture: Banking Infrastructure, International Operations

At a major international bank, routing inter-subsidiary SWIFT messages through an internally architected switch rather than the commercial SWIFT Gateway would eliminate per-transaction costs across its global subsidiary network. The Payment Switch initiative required two technically demanding deliverables: a rule-based routing engine and a complete rewrite of the Message Handler module — the component responsible for all incoming and outgoing message translations across the communication layer.

The governance challenge in this engagement was the risk surface of the implementation. The Message Handler was embedded at the core of all outgoing and incoming message translations for the entire communication system — a high-fragility component where a defect had institution-wide consequences. Module 4's acceptance criteria for mission-critical systems were applied throughout the development of the Payment Switch: standardized instrumentation from the initial build, explicit reliability targets (processing throughput, message translation accuracy, failure mode behavior under load), and CAPA processes for every defect identified in pre-production testing.

The outcome: the Payment Switch processed up to 100,000 SWIFT messages per hour in production, delivered a 30% reduction in operational communications costs, and achieved zero critical defects after production deployment.

### 4.4 Aggregate Outcomes

Across five large-scale modernization initiatives, aggregate outcomes include: a 30% reduction in development effort; a 35% reduction in testing cycles; a 25% reduction in time-to-market through governed GenAI-assisted acceleration; and more than USD 10 million in total revenue outcomes.

## 5. Policy Alignment: EMRGF as an Operational Implementation Layer for U.S. National Mandates

A distinctive characteristic of EMRGF is that it operationalizes governance expectations that U.S. national policy frameworks have established but not implemented at the enterprise level.

**NIST Cybersecurity Framework 2.0 (NIST CSF 2.0)** [2] defines six governance functions — Govern, Identify, Protect, Detect, Respond, Recover — that organizations are expected to institutionalize for cybersecurity risk management. It is a policy architecture: it defines what governance should achieve but does not provide the operational routines, change-control standards, or evidence artifacts that translate those expectations into daily practice. EMRGF's Cloud and Legacy Modernization Governance Pack fills this operational layer: architecture decision records, change-control gates, and audit-ready documentation standards operationalize the CSF 2.0 Govern and Protect functions at the migration level.

**NIST AI Risk Management Framework 1.0 (NIST AI RMF)** [3] and **Executive Order 14110** [5] establish national standards for governing AI risk through MAP, MEASURE, MANAGE, and GOVERN functions. Like CSF 2.0, the AI RMF does not prescribe how an engineering team should introduce GenAI-assisted workflows with auditable, controlled outputs. EMRGF's AI-Enabled Automation Governance Pack operationalizes AI RMF at this level: use-case scoped guardrails, validation checkpoints, output-review requirements, and controlled update gates translate governance expectations into daily engineering practice.

**Executive Order 14028** [4] mandates secure-by-design execution and controlled change across information systems. EMRGF's Security and Governance Gating tool operationalizes this through Module 1's change-control and exception-governance rules and Module 4's instrumentation and acceptance criteria — embedding security and reliability requirements in architecture decisions rather than applying them retroactively.

**FedRAMP** [11] addresses cloud service authorization — whether a service meets federal security standards — but not post-authorization governance: how an organization manages workloads, architecture decisions, data flows, and reliability controls within an authorized environment over time. EMRGF's Cloud Governance Pack and Measurement-by-Design tool fill this operating gap.

U.S. federal agencies and regulated enterprises face compliance mandates — from NIST, FedRAMP, and executive order — that define governance expectations they cannot currently meet through any standardized operational model. EMRGF provides that model, expressed through portable governance artifacts and routines adoptable across heterogeneous environments.

## 6. Deployment Model: Institutional Adoption Without Ongoing External Dependency

A governance framework that requires continuous external expertise to operate has not solved the governance problem — it has externalized it. EMRGF is specifically designed to become an internally owned institutional operating system through a copy-with-parameters deployment model.

**Copy-with-parameters** means that the same core governance routines, artifacts, control dictionaries, audit schedules, governance calendars, and training packages are versioned to reflect the specific constraints of each adopting organization: its regulatory exposure, its legacy footprint, its data criticality, its risk profile, and its existing technical stack. The governance logic is identical; the parameters that configure it are organization-specific. This model enables replication without reinvention — each adopting organization configures EMRGF to its environment rather than designing governance from scratch.

The **Training-of-Trainers Institutionalization Model** is the mechanism that ensures EMRGF survives the initial implementation. Internal stakeholders — IT leadership, cloud and platform engineering leads, data engineering and analytics leads, security and governance functions — are trained not just to apply EMRGF's routines but to own, sustain, teach, and evolve them. These internal multipliers become the organizational carrier of the framework, embedding it into the governance calendar through recurring audit cycles, quarterly review gates, and periodic revalidation routines that persist independently of external involvement.

EMRGF's highest-impact application sectors are financial services (where downtime and data defects generate cascading counterparty risk and regulatory sanction), healthcare (where EHR modernization failures directly affect patient safety and care continuity), industrial manufacturing (where supply chain systems depend on reliable data infrastructure), and federal IT (where the GAO documents persistent governance failures across defense, transportation, veterans services, and education [1]).

## 7. Discussion: Limitations and Open Problems

EMRGF's empirical results are drawn from single-site deployments where one architect applied the framework's governance principles in specific organizational contexts. Multi-organization validation studies — with independent practitioners applying EMRGF across different industries, governance maturity levels, and technical environments — are needed to establish generalizability. The outcome metrics reported in Section 4 are measured against project-specific baselines; common measurement instruments and baseline definitions would enable cross-organization comparison and strengthen the causal attribution.

The AI-Enabled Automation Governance Pack requires periodic revalidation — at minimum annually — as the GenAI capability landscape and regulatory expectations evolve. The training-of-trainers institutionalization model similarly requires standardized competency assessment to ensure high-fidelity knowledge transfer across personnel transitions. How EMRGF integrates with or partially replaces existing ITIL, COBIT, or SAFe structures in practice is an open research question that field adoption studies will need to address.

## 8. Conclusion

Enterprise technology modernization fails at an organizational level because governance fails, not because engineering fails. The evidence of this failure — documented by the GAO at $40 billion per year in recoverable federal IT waste alone — establishes that this is a systemic problem requiring a governance solution. Existing frameworks address adjacent concerns: ITIL manages service stability, COBIT defines governance objectives, TOGAF designs target architectures, and cloud provider well-architected frameworks guide platform-specific design decisions. None of them provides an integrated, portable, institutionalizable operating model for governing the full scope of controlled modernization — across migrations, data platforms, AI automation, and reliability under load.

EMRGF fills this gap. Its four modules — Cloud and Legacy Modernization Governance, Data Platform Reliability and Evidence Integrity, AI-Enabled Automation Governance, and Mission-Critical Reliability and Root-Cause Routines — address the specific failure modes that drive modernization governance failures. Its five implementation tools translate these modules into an adoptable institutional operating system. Its training-of-trainers model enables organizations to own the framework independently, without ongoing external dependency. And its explicit alignment with NIST CSF 2.0, NIST AI RMF, and Executive Orders 14028 and 14110 positions it as an operational implementation layer for mandates that currently lack one.

The practical takeaway for software architects, engineering leaders, and CIOs is direct: governance infrastructure is a prerequisite for modernization success, not an overhead to be minimized. Organizations that institutionalize controlled, evidence-based governance routines before initiating large-scale platform migrations — not after the first production incident — substantially reduce the cost, duration, and risk of modernization at scale. EMRGF provides the structured operating model for doing so.

## Author Biography

**Harveen Punihani** is a Senior Technical Architect at Impetus Technologies Inc. and an IEEE Senior Member. He has 24 years of experience delivering enterprise technology modernization, cloud platform engineering, data reliability, and AI-enabled automation governance across financial services, industrial manufacturing, and retail. He holds an M.Tech. from Indian Institute of Technology Kanpur and a B.Tech. from National Institute of Technology, and holds AWS Certified Solutions Architect, Google Cloud Professional Cloud Architect certifications.